\documentclass[aps,twocolumn,prd,superscriptaddress,nofootinbib]{revtex4}

\usepackage{amsmath}
\usepackage{graphicx}

\newcommand{\dslash}{\partial\hspace{-1.2ex}\slash}
\newcommand{\pslash}{p\hspace{-1ex}\slash}
\newcommand{\Tr}{\mathrm{Tr}}
\newcommand{\dfourp}{\frac{d^4 p}{(2 \pi)^4}}
\newcommand{\unity}{1\hspace{-1.3mm}1}
\newcommand{\eq}[1]{Eq.~(\ref{#1})}
\newcommand{\eqs}[1]{Eqs.~(\ref{#1})}

\begin{document}

\title{Pseudoscalar bosonic excitations in the
  color-flavor locked phase at moderate densities}

\author{Verena Kleinhaus}
\affiliation{Institut f\"ur Kernphysik, Technische Universit\"at Darmstadt, Germany}

\author{Michael Buballa}
\affiliation{Institut f\"ur Kernphysik, Technische Universit\"at Darmstadt, Germany}

\date{\today}

\begin{abstract}
The properties of pseudoscalar bosonic excitations in the color-flavor locked 
phase at moderate densities are studied within a model of the 
Nambu--Jona-Lasinio type. 
Our previous analysis \cite{Kleinhaus:2007ve} is extended to Goldstone
bosons with hidden flavor and to higher-lying modes which stay massive
in the chiral limit. 
The bosons are constructed explicitly by solving the Bethe-Salpeter 
equation for quark-quark scattering in random phase approximation.
The masses and weak decay constants of the Goldstone bosons are found
in good agreement with predictions from the low-energy effective theory.
In the non-Goldstone sector we find an $SU(3)$ octet which is weakly
bound, while the singlet appears to be unbound.
\end{abstract}

\maketitle


\section{Introduction}

The ground state of quark matter at asymptotically high baryon densities is
the color-flavor locked (CFL) phase \cite{Alford:1998mk}. In this phase up,
down, and strange quarks are paired in a particularly symmetric way. 
As all quark flavors and colors participate in a condensate, all
fermionic modes are gapped and do not appear in the low-energy excitation
spectrum. 
Spontaneous breaking of baryon number and chiral symmetry leads to the 
emergence of one scalar and eight pseudoscalar Goldstone bosons. 
A ninth pseudoscalar Goldstone boson is expected due 
to the spontaneous breaking of the $U_A(1)$ symmetry, which is a symmetry
of QCD at high density \cite{Schafer:2002ty,Rapp:1999qa}.
These Goldstone bosons are the lowest lying excitations and should
play an important role for the thermodynamics of strongly interacting 
matter at high density. In nature, this could have phenomenological 
consequences for compact star physics
\cite{Shovkovy:2002kv,Manuel:2004iv,Alford:2007rw,Alford:2008pb}.

The pseudoscalar Goldstone bosons in the CFL phase have been studied
extensively within the low-energy effective theory (LEET) which is based 
on the symmetry breaking pattern
\cite{Casalbuoni:1999wu,Son:1999cm,Schafer:2000ew,
Casalbuoni:2000na,Bedaque:2001je,PhysRevD.65.054042,Yamamoto:2007ah}.
At very high densities, the corresponding low-energy constants can be
calculated from QCD within high density effective theory (HDET) 
\cite{Bedaque:2001je,Hong:1998tn,Hong:1999ru,Beane:2000ms,Nardulli:2002ma}
. 
The bosons form an $SU(3)$ octet and a singlet and there is 
a one-to-one correspondence to the well-known octet and singlet of 
pseudoscalar mesons in vacuum. 
They are therefore often called ``mesons'' as well and can be 
identified with pions, kaons, $\eta$, and $\eta'$, according to their quantum 
numbers. 
An interesting prediction made within the LEET is that the strange quark mass
acts as an effective strangeness chemical potential, eventually leading
to kaon condensation in the CFL phase
\cite{Schafer:2000ew,Bedaque:2001je,PhysRevD.65.054042}.

So far only few authors have investigated the CFL mesons starting from 
quark degrees of freedom \cite{Kleinhaus:2007ve,Rho:1999xf,Rho:2000ww,Ebert:2006tc,Ebert:2007bp}
In Ref.~\cite{Kleinhaus:2007ve} we have studied meson masses and decay 
constants as well as the onset of kaon condensation within an NJL-type model
at moderate densities. 
In the numerical part, we have restricted ourselves to open-flavor
Goldstone bosons, i.e., kaons and charged pions, 
which are technically easier to describe.
Moreover, this is the sector where meson condensation can occur. 
Our results are consistent with the LEET analysis, but we found
quantitative differences from the weak-coupling limit of QCD. 

One focus of the present article is to extend these studies
to the hidden-flavor sector, i.e., $\pi^0$, $\eta$, and $\eta'$. 
Again, we will compare our results with the predictions of the LEET. 

In spite of the formal correspondence of the pseudoscalar
Goldstone bosons in the CFL phase and in vacuum, their physical
nature is very different. In fact the ``mesons'' in the CFL phase
are mainly superpositions of diquark and di-hole states, rather
than quark-antiquark states. This is possible because baryon number
is not a good quantum number in the CFL phase.  
In Ref.~\cite{Kleinhaus:2007ve} we argued that there is 
a second octet and a second singlet of pseudoscalar bosons, which come 
about as superpositions of diquark and di-hole states orthogonal to the 
Goldstone modes. These higher-lying excitations have been investigated 
in Ref.~\cite{Ebert:2007bp}, but only in the chiral limit.
In the present paper, we study their properties for equal and non-equal
quark masses. 

This article is organized as follows.
In Sec.~\ref{formalism} we introduce our model and briefly summarize the 
formalism developed in Ref.~\cite{Kleinhaus:2007ve}.
After that we present our results for the Goldstone bosons in
Sec.~\ref{results:GB} and for the higher-lying modes in
Sec.~\ref{results:he}. We conclude with a short summary in 
Sec.~\ref{summary}.


\section{Formalism}
\label{formalism}

In this section we briefly summarize our NJL model for pseudoscalar
mesons in the CFL phase. Further details can be found in 
Ref.~\cite{Kleinhaus:2007ve}.

We consider the Lagrangian 
\begin{equation}
\mathcal{L} = \bar{q} (i \dslash - \hat{m}) q 
+  \mathcal{L}_\mathit{qq},
\label{eq:L}
\end{equation}
where $q$ is a quark field with three flavor and three color degrees 
of freedom, $\hat m=\mathrm{diag}_f(m_u,m_d,m_s)$ is the mass matrix, and
\begin{alignat}{1}
\mathcal{L}_\mathit{qq}
= H \hspace{-3mm} \sum_{A,A'=2,5,7} \big[ \quad 
                      & ( \bar{q} i \gamma_5 \tau_A \lambda_{A'} C\bar{q}^T) 
                      ( q^T C i \gamma_5 \tau_A \lambda_{A'} q) 
\nonumber\\
            +\;         & ( \bar{q} \tau_A \lambda_{A'} C \bar{q}^T)
                      ( q^T C \tau_A \lambda_{A'} q )\; \big]
\label{eq:Lqq}
\end{alignat}
describes an $SU(3)_\mathit{color} \times U(3)_L\times U(3)_R$
symmetric four-point interaction with a dimensionful coupling constant $H$.
$C=i \gamma^2 \gamma^0$ is the matrix of charge conjugation,
and $\tau$ and $\lambda$, denote Gell-Mann matrices acting in flavor and 
color space, respectively. 
In this article, we follow the convention that the indices 
$A$ and $A'$ are used for the antisymmetric Gell-Mann matrices only, i.e.,
$A,\, A' \in \{2,5,7\}$.

Working in Nambu-Gorkov formalism, the interaction Lagrangian
gives rise to the quark-quark scattering kernel
\begin{equation}
\hat K = \Gamma_i K_{ij} \bar\Gamma_j~,\qquad K_{ij} = 4H\delta_{ij}~,
\label{eq:K}
\end{equation}
where $\bar{\Gamma}_i=\gamma_0 \Gamma_i^\dagger \gamma_0$
with 18 scalar operators
\begin{equation}
 \Gamma_{AA'}^{s\uparrow} =
 \begin{pmatrix}
    	0 & i \gamma_5 \tau_A \lambda_{A'} \\
    	0 & 0
    \end{pmatrix} , \quad
 \Gamma_{AA'}^{s\downarrow} =
 \begin{pmatrix}
  	0 & 0 \\ 
 	i \gamma_5 \tau_A \lambda_{A'} & 0
 \end{pmatrix}
\label{eq.scalar_vertices}
\end{equation}
and 18 pseudoscalar operators
\begin{equation}
 \Gamma_{AA'}^{p\uparrow} =
 \begin{pmatrix}
    	0 & \tau_A \lambda_{A'} \\
    	0 & 0
 \end{pmatrix} , \quad
 \Gamma_{AA'}^{p\downarrow} =
 \begin{pmatrix}
 	0 & 0 \\ 
	\tau_A \lambda_{A'} & 0
 \end{pmatrix} \,. 
\label{eq.ps_vertices}
\end{equation}
In the CFL phase, the inverse dressed quark propagator in Nambu-Gorkov space
is given by
\begin{equation}
 S^{-1}(p) = \begin{pmatrix}
 	\pslash + \hat{\mu}\gamma^0 - \hat{m}  &  
        \sum\limits_{A=2,5,7}\Delta_A \gamma_5 \tau_A \lambda_A \\
        -\hspace{-3mm}\sum\limits_{A=2,5,7}\Delta_A^*\gamma_5\tau_A\lambda_A  &  
        \pslash - \hat{\mu}\gamma^0 - \hat{m}
	\end{pmatrix} \, .
\label{eq.Sinv}
\end{equation}
The gap parameters $\Delta_A$, which enter the anomalous components, are
obtained through minimizing the mean-field thermodynamic potential $\Omega$,
leading to three gap equations,
$\frac{\partial \Omega}{\partial \Delta_A^*}=0$.
In addition, we require electric and color neutrality, 
which fixes the electric and color chemical potentials at given temperature
and quark number chemical potential.
In the following we only consider the (fully gapped) CFL phase at zero 
temperature in the isospin symmetric limit, $m_u = m_d$.
In this case, we only need a nonzero color chemical potential $\mu_8$
to ensure neutrality.

We calculate the mesonic excitations by solving the RPA equation for the
$T$-matrix in Nambu-Gorkov space,
\begin{equation}
 	T(q) = K + K J(q) T(q) = [\unity-KJ(q)]^{-1}K\, ,
\label{eq:T}
\end{equation}
where the matrix $K$ is related to the scattering kernel, see \eq{eq:K}.
The elements of the one-loop polarization matrix $J(q)$ are given by
\begin{equation}
  J_{ij}(q) = i \int \frac{d^4 k}{(2 \pi)^4} \, 
  \text{Tr} \left[\bar{\Gamma}_i S(k+q) \Gamma_j
  S(k) \right] \,,
\label{eq:J_ij}
\end{equation}
where we have introduced a ``vacuum-like'' notation for brevity.  
In medium the zero components of $q$ and $k$ should be replaced by
bosonic and fermionic Matsubara frequencies, respectively, and we
should replace $i\int\frac{dk^0}{2\pi}$ by the Matsubara sum
$-T\sum_n$. The remaining three-momentum integral is divergent and will be
regularized by a sharp cutoff $\Lambda$.

The matrices $T$ and $J$ are $36\times 36$ matrices in the space of the
operators given in Eq.~(\ref{eq.scalar_vertices}) and
(\ref{eq.ps_vertices}). They are block diagonal because scalar and
pseudoscalar operators do not mix with each other. The two $18\times 18$
blocks can be decomposed further, each into six $2 \times 2$ blocks and one 
$6 \times 6$ block. 
These blocks can be attributed to different flavor quantum numbers:
The $2 \times 2$ blocks correspond to the open-flavor mesons
(i.e., in the pseudoscalar sector, $\pi^+$, $\pi^-$, $K^+$, $K^-$, $K^0$, 
$\bar K^0$), while the $6 \times 6$ block contains the hidden-flavor mesons
($\pi^0$, $\eta_8$, $\eta_0$). Here $\eta_8$ and $\eta_0$ are the $SU(3)$
octet and singlet states. As in vacuum they mix to the physical states
$\eta$ and $\eta'$ when flavor $SU(3)$ is broken explicitly.

By diagonalizing the blocks we get one Goldstone mode and one massive 
excitation for each meson. In the vicinity of their poles, we can 
parameterize each mode as a free boson with mass $m_M$ in the presence of 
a boson chemical potential $\mu_M$.
In the following, we restrict ourselves to the meson rest frame,
$q = (q_0, \vec 0)$. Then the parameterization reads
\begin{equation}
 T^{(M)}(q_0) \approx \frac{-g_M^2}{(q_0+\mu_M)^2-m_M^2}\, .
\label{eq:Ti}
\end{equation}
The constant $g_M$ is a wave function renormalization constant, which
can be interpreted as a coupling constant of the boson to an external 
quark.

The (time-like) weak decay constant $f_M$ of the pseudoscalar meson $M$ 
is given by the expression
\begin{equation}
  f_M = \left. \frac{1}{q_0}\!\int\! \dfourp \frac{1}{2}
  \Tr [\bar{A}_M^0 S(p+q) g_M \Gamma_M' S(p)]\right|_{
  \begin{array}{l}
  \scriptstyle q_0 = m_M,\\ 
   \scriptstyle \vec q=0
  \end{array}
  }.
\label{eq:fi}
\end{equation}
Here $\Gamma_M'$ denotes the meson-quark-quark vertex of the eigenmode $M$
resulting from the diagonalization discussed above.  
$\bar{A}_M^0$ is the vertex of an external axial current in the 
corresponding flavor channel. 

In this paper we evaluate the decay constants in the flavor-$SU(3)$ limit
only. 
We then only need the singlet channel, i.e., the $\eta^0$ and
one representative of the octet, e.g., the $\pi^0$.
The corresponding vertices of the external axial currents read 
\begin{equation}
  \bar{A}^0_{\eta_0} = \gamma^0 \gamma_5 \frac{\tau_0}{4}\; \unity_{NG} 
\label{eq:A0sing}
\end{equation}
and
\begin{equation}
  \bar{A}^0_{\pi^0} = \gamma^0 \gamma_5 \frac{\tau_3}{2}\; \unity_{NG}~, 
\label{eq:A0oct}
\end{equation}
where $\unity_{NG}$ indicates a unit matrix in Nambu-Gorkov space. 
$\tau_0 = \sqrt{\frac{2}{3}}\unity_f$ is proportional to the unit
matrix in flavor space.\footnote{
The exact normalization of the generators in flavor space is a matter
of convention. The normalization conventions in \eqs{eq:A0sing} and 
(\ref{eq:A0oct}) have been chosen to be consistent with the standard
definitions of the weak decay constants in the LEET approach.
Note that the generator in the singlet channel, $t_0 = \frac{\tau_0}{4}$, 
is a factor of two smaller than introduced in our previous paper 
\cite{Kleinhaus:2007ve}.}


\section{Goldstone bosons}
\label{results:GB}

In this section we discuss our numerical results for the pseudoscalar
Goldstone bosons. The results for the higher-lying pseudoscalar excitations 
will be presented in Sec.~\ref{results:he}.

We restrict ourselves to $T=0$ and a fixed quark number chemical potential
$\mu=500$~MeV. We choose a three-momentum cuttoff $\Lambda=600$~MeV and, 
if not otherwise stated, a diquark coupling $H=1.4\,\Lambda^{-2}$. 
For these parameters we have $\Delta = 79.1$~MeV in the chiral limit
and we are in the fully gapped CFL phase for all values of $m_s$ we
consider.

\subsection{Weak decay constants in the chiral limit}
\label{results:GB:fpi}

We begin with the weak decay constants in the chiral limit,
$m_u=m_d=m_s=0$. 
In the chiral limit all nine pseudoscalar Goldstone bosons are exactly
massless. However, since the CFL ground state is symmetric under a residual 
$SU(3)$ symmetry, but not under $U(1)$, we have to distinguish between 
the decay constants in the octet (``$f_\pi$'') and in the singlet 
(``$f_{\eta'}$'').  


\begin{figure}[hbt]
\begin{center}
 \includegraphics[width=\linewidth]{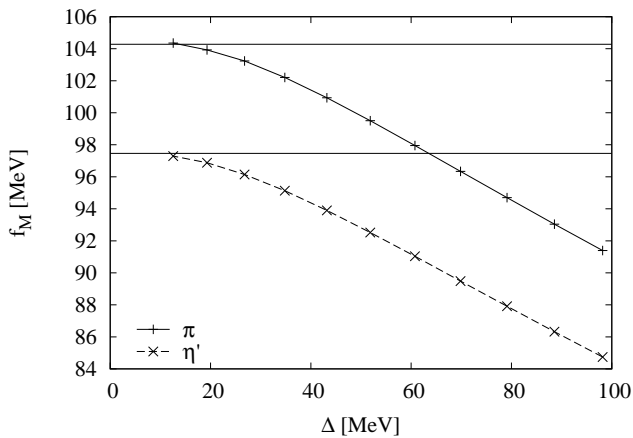}
 \caption{Weak decay constants $f_\pi$ and $f_{\eta'}$ in the chiral limit
  as functions of the gap parameter $\Delta$.
  Thin horizontal lines: weak-coupling limit.}
\label{fig:f_pi}
\end{center}
\end{figure}


Our numerical results are shown in Fig.~\ref{fig:f_pi}.
For technical reasons the calculations have been performed with  
$m_u=m_d=m_s=0.1$~MeV, but the difference to the exact chiral limit 
is negligible. 
In the figure, $f_\pi$ and $f_{\eta'}$ are displayed as functions of 
the gap parameter $\Delta$, which was varied by varying the coupling 
constant $H$. 
The numerical points are indicated by the points, which have been
connected by straight lines to guide the eye. 
The curve for the octet decay constant $f_\pi$ was already shown in
Ref.~\cite{Kleinhaus:2007ve}. There we restricted ourselves to
mesons with open flavor, but the decay constants of the hidden-flavor 
octet mesons, $\pi^0$ and $\eta$, are of course the same.
The decay constant in the singlet channel, $f_{\eta'}$, was not calculated
in Ref.~\cite{Kleinhaus:2007ve}. As one can see in the figure, it is
a few percent larger than $f_\pi$. 
This is well understood in the weak-coupling limit, where $f_\pi$ and 
$f_{\eta'}$ are given by \cite{Son:1999cm}
\begin{equation}
	f_{\pi}^2 = \frac{21-8 \ln(2)}{18}\left(\frac{\mu^2}{2\pi^2}\right) 
\quad \text{and} \quad
	f_{\eta'}^2 = \frac{3}{4} \left(\frac{\mu^2}{2\pi^2}\right) \,.
\label{eq:fpietapweak}
\end{equation}
These limiting values are marked by the thin horizontal lines in 
Fig.~\ref{fig:f_pi}.
Indeed, for $\Delta \rightarrow 0$ our results approach the 
weak-coupling limit, both, in the singlet and in the octet channel. 
For $f_\pi$, this was confirmed in a more careful analysis in 
Ref.~\cite{Kleinhaus:2007ve}, where a semi-analytical formula for the
decay constant was derived. An analogous analysis could be done for 
$f_{\eta'}$ as well. 

For larger values of $\Delta$, we find significant deviations from 
\eq{eq:fpietapweak}.
This is not surprising as the weak-coupling assumption is not
valid in that region.


\subsection{Meson masses}

Next, we discuss the masses of the (pseudo) Goldstone bosons for non-vanishing 
quark masses. 
As pointed out earlier, our focus is on the hidden-flavor sector.
To obtain a complete picture, however,
we will also include our results for the open-flavor
mesons, which have been discussed already in great detail in
Ref.~\cite{Kleinhaus:2007ve}.


\begin{figure}[hbt]
\begin{center}
 \includegraphics[width=\linewidth]{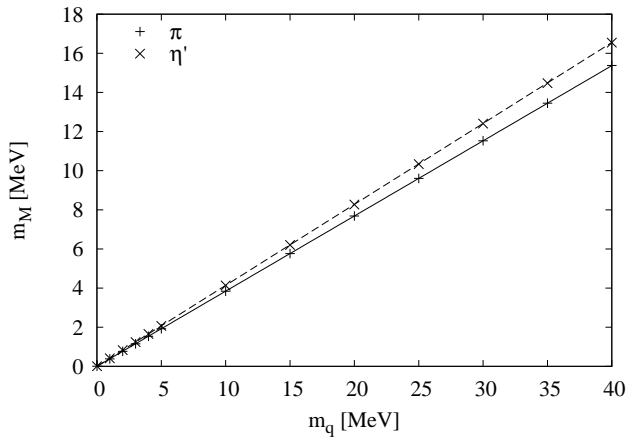}
 \caption{Masses of the pseudoscalar Goldstone bosons
          as functions of a common quark mass. 
 Points: numerical results, lines: LEET prediction.}
\label{fig:H1.4_hidden}
\end{center}
\end{figure}


We begin with the case of non-zero, but equal quark masses 
$m_u=m_d=m_s \equiv m_q$. 
Our results are shown in Fig.~\ref{fig:H1.4_hidden},
where the numerical calculations are indicated by the points.
Since the $SU(3)$ symmetry is still intact, the octet remains degenerate.
This means, $\pi^0$ and $\eta \equiv \eta_8$ have the same mass $m_\pi$
as the flavored mesons discussed in Ref.~\cite{Kleinhaus:2007ve}.
The $\eta' \equiv \eta_0$, on the other hand, forms a singlet and 
has a slightly higher mass for $m_q \neq 0$.

To very high accuracy, both, the octet and the singlet masses, grow
linearly with the quark mass. This is expected from the LEET, where,
to leading order in the quark mass,
the octet \cite{Bedaque:2001je} and the singlet \cite{Schafer:2001za}
masses are given by 
\begin{equation}
	m_\pi = \sqrt{\frac{8A}{f_\pi^2}} m_q
\quad \text{and} \quad 
	m_{\eta'} = \sqrt{\frac{8A}{f_{\eta'}^2}} m_q~,
	\label{eq:m_octetsinglet}
\end{equation}
respectively.
Whereas this form is universal, the constant $A$, which determines
the coefficient, depends on the interaction. For QCD in the weak-coupling
limit it has been calculated in Ref.~\cite{Son:1999cm}.
In Ref.~\cite{Kleinhaus:2007ve} we derived an analytical expression for 
$A$ for the present NJL model (see Eq.~(E1) in \cite{Kleinhaus:2007ve}).
Inserting that formula into \eq{eq:m_octetsinglet} and
using the chiral limit values of
$f_\pi$ and $f_{\eta'}$ from the previous subsection, we obtain the
results which are indicated by the lines in Fig.~\ref{fig:H1.4_hidden}.
Obviously, they are in excellent agreement with the numerical calculations.

\begin{figure}[htb]
\begin{center}
 \includegraphics[width=\linewidth]{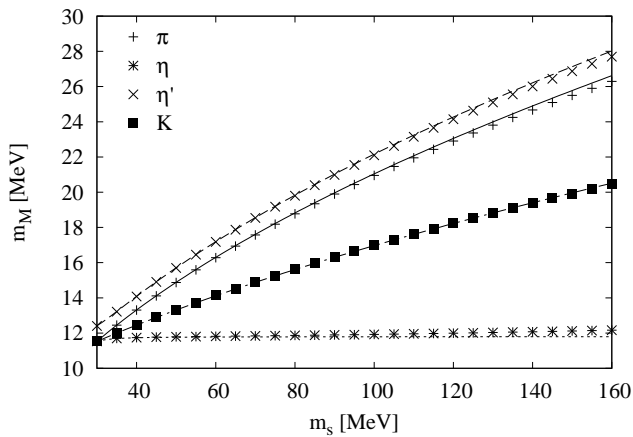}
 \caption{Masses of the pseudoscalar Goldstone bosons as functions of
 the strange quark mass $m_s$ for $m_u=m_d=30$\,MeV.
 Points: numerical results, lines: LEET fit.}
\label{fig:m30_hidden}
\end{center}
\end{figure}

Next, we study the effect of unequal quark masses. We choose a fixed
mass of 30\,MeV for $m_u$ and $m_d$ and vary the strange quark mass. 
The results are shown in Fig.~\ref{fig:m30_hidden}.
Since the hidden-flavor mesons are not sensitive to the effective
strangeness chemical potential induced by $m_s-m_u$, their masses are
directly given by the poles of the $T$-matrix.
For comparison we also display the kaon mass in the figure,
which has already been shown in Ref.~\cite{Kleinhaus:2007ve}.


Our numerical results are indicated by the points.
Because of the explicit breaking of the $SU(3)$-flavor symmetry, 
the octet is no longer degenerate. On the other hand, since we kept
$m_u = m_d \equiv m_q$, isospin is still an exact symmetry of the Lagrangian. 
Therefore, the $\pi^0$ has the same mass as the charged pions
already discussed in  Ref.~\cite{Kleinhaus:2007ve}.

The masses of $\pi$, $K$ and $\eta$ exhibit the ``inverse ordering''
already predicted in Ref.~\cite{Son:1999cm},
with the pion being the heaviest and the $\eta$ being the lightest
among these excitations.
In fact, the mass of the $\eta$ meson barely changes with increasing 
strange quark mass. The $\eta'$, on the other hand, drops out of the 
inverse ordering scheme and is slightly heavier than the pion. 

To understand this behavior,
we again compare our results with the LEET predictions.
For $\pi$ and $K$ these are given by
\cite{Son:1999cm,Bedaque:2001je,PhysRevD.65.054042,Schafer:2001za}
\begin{equation}
   m_{\pi}^2 = \frac{8A}{f_\pi^2} m_s m_q~, 
   \quad \text{and} \quad
   m_{K}^2 = \frac{4A}{f_\pi^2} m_q (m_q+m_s)~.
\label{eq:m_piK}
\end{equation}
For $\eta$ and $\eta'$ the situation is more complicated because of the
mixing of the octet with the singlet for $m_s \neq m_q$. To that end we
have to diagonalize the mass matrix
\begin{equation}
 \hat M^2 =  
\begin{pmatrix}
 	m_{\eta_0}^2 & m_\mathit{mix}^2 \\ 
	m_\mathit{mix}^2 & m_{\eta_8}^2
 \end{pmatrix} \, ,
\label{eq:m_matrix}
\end{equation}
where 
\begin{equation}
	m_{\eta_0}^2 = \frac{8A}{3f_{\eta'}^2} m_q(2 m_s+m_q), \quad
	m_{\eta_8}^2 = \frac{8A}{3 f_\pi^2} m_q (m_s + 2 m_q),
\label{eq:m_eta08}
\end{equation}
are the singlet and octet masses, respectively, and
\begin{equation}
   m_\mathit{mix}^2 = \frac{8\sqrt{2}A}{3 f_{\eta'} f_\pi} m_q (m_s - m_q)
\label{eq:m_mix}
\end{equation}
describes the mixing \cite{Schafer:2001za}.

Inserting the values for $f_\pi$, $f_{\eta'}$, and $A$
used in the equal mass case into \eqs{eq:m_piK}, (\ref{eq:m_eta08})
and (\ref{eq:m_mix}), and diagonalizing \eq{eq:m_matrix}, we obtain
the mass eigenvalues which are indicated by  the lines in 
Fig.~\ref{fig:m30_hidden}. Obviously, the agreement with the numerical
NJL results is very good, with small deviations only at higher values
of $m_s$. 

To get more compact expressions for the $\eta$ and $\eta'$ masses,
we can make use of the fact that the difference between
$f_\pi$ and $f_{\eta'}$ is small.
We may thus write
\begin{equation}
    f_{\eta'}^2 = f_\pi^2 - \delta f^2
\end{equation}
and expand the eigenvalues of the mass matrix \eq{eq:m_matrix}
until the order $\delta f^2$.
One finds
\begin{equation}
  m_\eta^2 \approx \frac{8A}{f_\pi^2}\, (1 + \frac{1}{3}\frac{\delta f^2}
 {f_\pi^2}) \, m_q^2
 \phantom{ m_s = (1 + \frac{2}{3}\frac{\delta f^2}{f_\pi^2}) \, m_\pi^2~.}
\end{equation}
and
\begin{equation}
  m_{\eta'}^2 \approx \frac{8A}{f_\pi^2}\, (1 + \frac{2}{3}\frac{\delta f^2}
 {f_\pi^2}) \, m_q m_s = 
 (1 + \frac{2}{3}\frac{\delta f^2}{f_\pi^2}) \, m_\pi^2~.
\end{equation}
This explains why the $\eta$ mass stays (approximately) constant with
$m_s$, whereas the $\eta'$ behaves very similar to the pion, but scaled 
by a constant factor.

\section{Higher-lying excitations}
\label{results:he}

As explained in Sec.~\ref{formalism}, besides the Goldstone modes, there 
also exist an octet and a singlet of higher-lying pseudoscalar excitations, 
which stay massive even in the chiral limit. In this section, we analyze
the masses and decay constants of these mesons.

\subsection{Masses}

In Ref.~\cite{Ebert:2007bp} the higher-lying pseudoscalar modes in the CFL 
phase have been studied in the chiral limit within a similar model.
The authors did not find any singlet solutions, 
whereas for the octet they report the existence of resonance states above
$2\Delta$, i.e., the threshold for decay into two quasiparticles.
These solutions were identified as poles on the second Riemann sheet
in the complex energy plane. 

\begin{figure}[hbt]
\begin{center}
 \includegraphics[width=\linewidth]{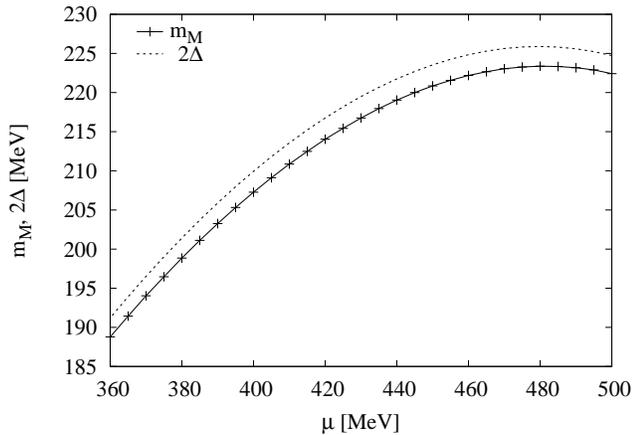}
 \caption{The higher-lying pseudoscalar octet mode in the chiral limit
  as a function of the quark chemical potential $\mu$.
  The parameters are $H=1.73925\,\Lambda^{-2}$, $\Lambda=602.3$\,MeV
  \cite{Ebert:2007bp}.
  The dotted line indicates the decay threshold $2\Delta$.}
\label{fig:massive_mu}
\end{center}
\end{figure}

The model of Ref.~\cite{Ebert:2007bp} is practically the same as  
ours.\footnote{The Lagrangian used in Ref.~\cite{Ebert:2007bp} contains an 
additional quark-antiquark interaction. 
However, in the CFL phase in the chiral limit this term does not contribute
to the modes we discuss here.} 
It was thus to our surprise that we found bound-state solutions in the 
higher-lying octet.
This is shown in Fig.~\ref{fig:massive_mu}.
For a direct comparison with Ref.~\cite{Ebert:2007bp}, we used the 
same parameters ($\Lambda=602.3$\,MeV and $H=1.73925\,\Lambda^{-2}$)
and calculated the octet mass as a function of the quark number 
chemical potential $\mu$ (solid line).
In the chosen interval, 360~MeV~$< \mu <500$\,MeV, the masses of the 
excitations vary between 189 and 223\,MeV.
We also show the decay-threshold $2 \Delta$ (dotted line). 
As one can see, the meson masses closely follow
this line, but always stay below. This means that the octet modes are
bound in the entire interval. 


For technical reasons, the chiral limit was again approximated in our 
code by taking very small quark masses, $m_u=m_d=m_s=0.1$~MeV.
On the other hand, the authors of Ref.~\cite{Ebert:2007bp} restricted 
themselves to the chiral limit from the beginning. As a consequence they 
were able to derive a somewhat simpler equation for the meson masses,
see Eq.~(33) in \cite{Ebert:2007bp}. 
We checked that this equation has indeed bound-state solutions, 
which are in excellent agreement with the results shown in 
Fig.~\ref{fig:massive_mu}.
We are therefore convinced that our results (as well as 
Eq.~(33) in \cite{Ebert:2007bp}) are correct. 

We would like to point out that it is not excluded that there are
several branches of solutions. This means, we cannot exclude that the 
resonance-state solutions found in Ref.~\cite{Ebert:2007bp} are correct
as well. Unfortunately, 
the extension of our method to evaluate the polarization integral \eq{eq:J_ij}
above threshold would require additional effort, which is beyond the scope 
of the present paper. Our analysis is therefore restricted to the regime below
threshold.
For the same reason we could not determine the mass of the higher-lying
excitation in the singlet channel, which is always above 
threshold.\footnote{The behavior of the $T$-matrix below threshold
seems to indicate that the mass of the singlet meson is rather close to
the threshold as well. However, because of threshold effects a quantitative
estimate is difficult.}


\begin{figure}[hbt]
\begin{center}
 \includegraphics[width=\linewidth]{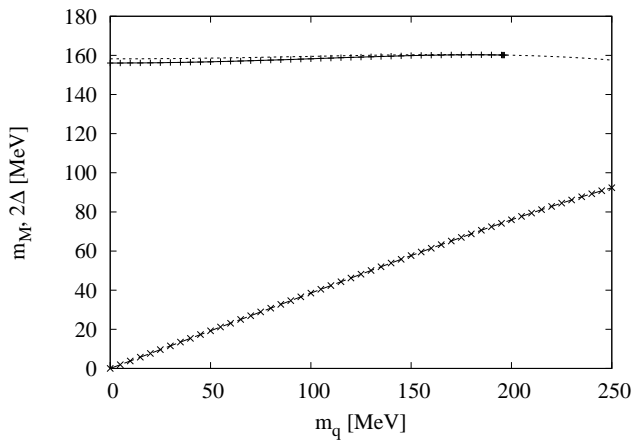}
 \caption{Masses of the pseudoscalar octet excitations
as functions of a common quark mass $m_q$: higher-lying modes
in comparison with the Goldstone modes. 
The dotted line indicates the decay threshold $2\Delta$.
}
\label{fig:masses_H1.4}
\end{center}
\end{figure}


We now return to our standard parameters 
($\mu=500$~MeV, $\Lambda=600$~MeV, $H=1.4\,\Lambda^{-2}$) 
and discuss the effect of finite quark masses.
As before, we begin with the simplified case of equal quark masses,
$m_u=m_d=m_s\equiv m_q$. 
In Fig.~\ref{fig:masses_H1.4} we show the masses of the two octets 
(Goldstone bosons and higher-lying excitations) as functions of $m_q$. 
While the masses of the Goldstone bosons increase linearly with a
sizeable slope, the masses of the higher-lying excitations stay nearly 
constant. 
In fact, they stay again very close to the decay threshold
$2\Delta$ (dotted line).
With increasing quark mass they approach this threshold and
the mesons become eventually unbound at $m_q \approx 190$~MeV.
At this point our curve terminates, because, as mentioned above,
our method does not allow to find solutions above threshold.

Next, we study the effect of an explicit breaking of the 
$SU(3)$-flavor symmetry on the higher-lying excitations. 
Again, we choose $m_u=m_d=30$\,MeV and vary the strange quark mass $m_s$.

\begin{figure}[hbt]
\begin{center}
 \includegraphics[width=\linewidth]{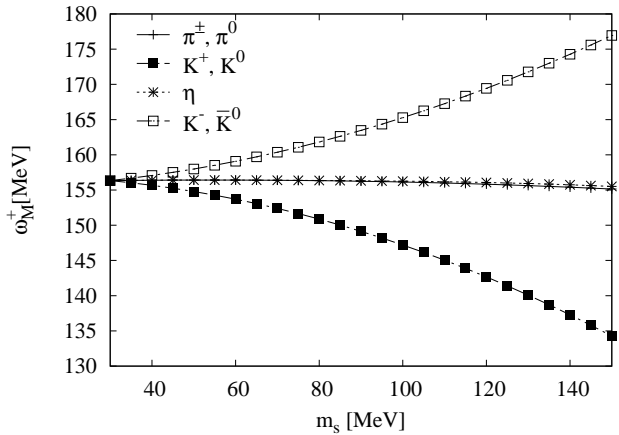}
 \caption{Positive pole positions of the $T$-matrix for the
 higher-lying modes as a function of the strange quark 
 mass $m_s$ for $m_u=m_d=30$\,MeV.}
\label{fig:m30_massive}
\end{center}
\end{figure}

As in the case of equal quark masses, we only find higher-lying 
excitations for the
octet modes. The positions of the poles are plotted in
Fig.~\ref{fig:m30_massive}. The poles of the pions\footnote{
The names ``pions'', ``kaons'', etc. refer again to the flavor
quantum numbers of the excitations.
We will not introduce new names for the higher-lying modes
as long as a confusion with the corresponding Goldstone modes
can be excluded.}and the $\eta$ stay nearly
constant around 155\,MeV whereas the pole of the antikaons ($K^-$ and
$\bar{K}^0$) grows from 156\,MeV for a quark mass of 30\,MeV to 177\,MeV for a
quark mass of 140\,MeV. For the kaons ($K^+$ and $K^0$) the situation is
reversed, the position of their pole moves from 156\,MeV to 
136\,MeV.


It should be noted that in the case of unequal quark masses
the threshold for decay into two quasiparticles is no longer
equal to $2 \Delta$. 
In fact, in this case the fermionic excitation spectrum contains
five different particle branches (two singlets, two doublets
and one triplet) with different excitation gaps, see, e.g.
Ref.~\cite{Abuki:2005ms}.
Thus, the decay threshold for the different meson modes depends
in a complicated way on their respective quasiparticle composition
and rises with $m_s$ in some channels, while it drops in others. 
It turns out that the splitting of the various mesonic modes
closely follows the splitting of the decay thresholds. 

\begin{figure}[hbt]
\begin{center}
 \includegraphics[width=\linewidth]{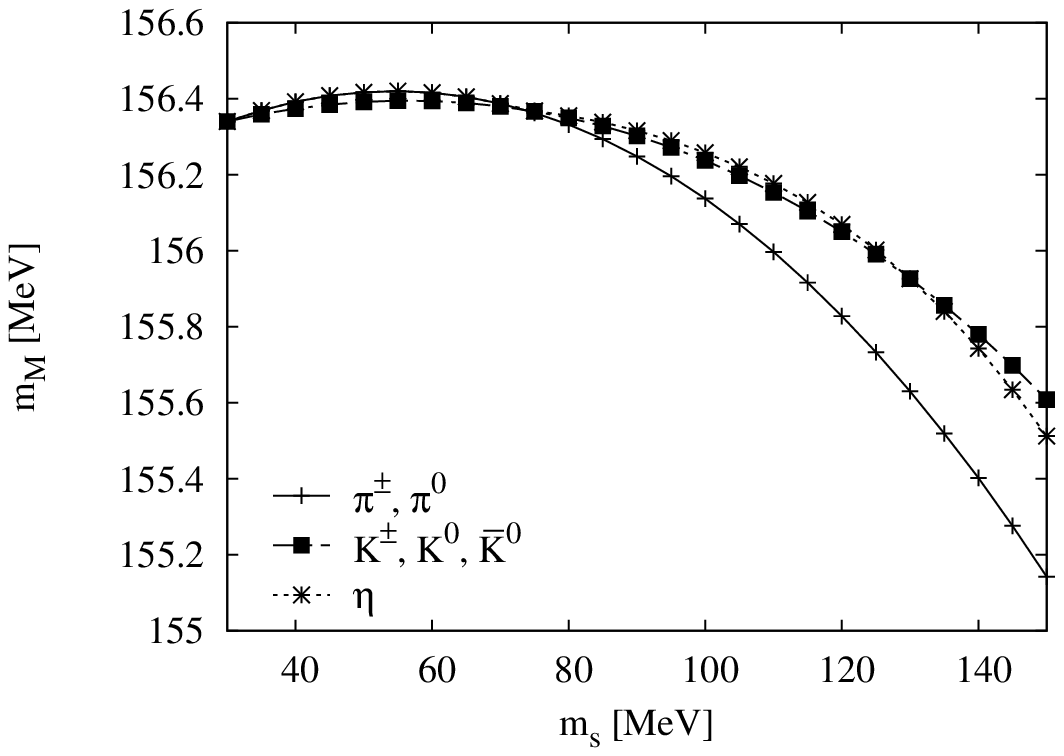}
 \includegraphics[width=\linewidth]{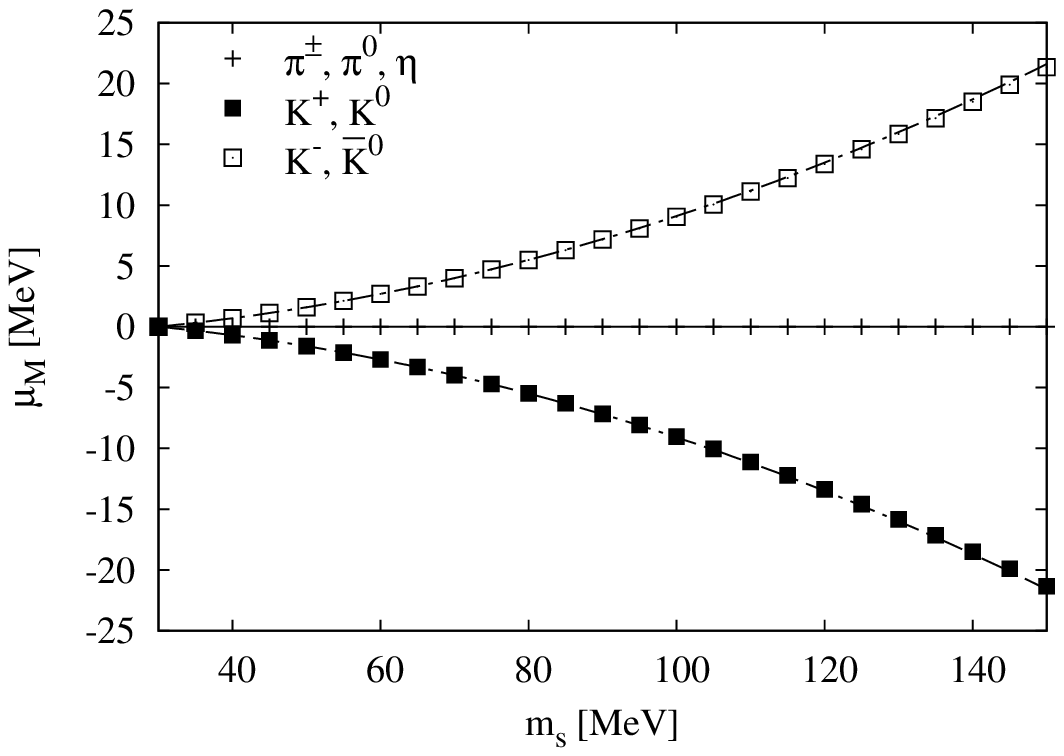}
 \caption{Masses (upper panel) and effective meson chemical potentials 
  (lower panel) of the higher-lying excitations
  as functions of the strange quark mass $m_s$ for $m_u=m_d=30$\,MeV.
  The various points indicate the numerical calculations using
  \eq{eq:mmu_M}. In the upper panel, these points have been connected
  by straight lines to guide the eye. 
  The lines in the lower part correspond to Eq.~(\ref{eq:chem_pot}).}
\label{fig:m30_massive_masses}
\end{center}
\end{figure}

In Fig.~\ref{fig:m30_massive} we only show the pole positions 
$\omega_M^+$ at positive values of $q_0$. In addition,
each mode has another pole $\omega_M^-$ at negative energies,
which is not shown in the figure.
Applying \eq{eq:Ti} we can calculate the masses and
effective chemical potential for each meson as
\begin{equation}
	m_M=\frac{1}{2}(\omega_M^+ - \omega_M^-), \quad 
        \mu_M=-\frac{1}{2}(\omega_M^+ + \omega_M^-).
\label{eq:mmu_M}
\end{equation}
The results are plotted in Fig.~\ref{fig:m30_massive_masses}. 
In the upper part of the figure the masses are shown.
They first slightly increase and then
decrease with $m_s$. However, this happens on a very small scale from 156.4 to
155.2\,MeV, i.e., unlike the Goldstone bosons, they stay nearly
constant. 
On this scale we can also see that the $\eta$ and the pions are not 
degenerate. (This was already the case for the poles shown in 
Fig.~\ref{fig:m30_massive}, but hardly visible.)


In the lower part of Fig.~\ref{fig:m30_massive_masses} we show
the effective meson chemical potentials.
Our numerical results are indicated by the points.
The chemical potentials vanish identically in the pion and $\eta$ channels,
whereas they are negative for kaons and positive for antikaons, with
equal absolute values.

Of course, this was to be expected: 
Since the ``effective meson chemical potentials'' are induced by
the mass differences of different quark flavors, and since we kept
$m_u=m_d$, there is only an effective strangeness chemical potential,
but no isospin chemical potential. 
Hence, pions and $\eta$ remain unaffected, while kaons and antikaons
feel opposite chemical potentials.
In fact, the effective meson chemical potentials should only depend on
the flavor content of the mesons. This means, there should be no 
difference between Goldstone bosons and higher-lying modes. Therefore,
we can compare our numerical results with the LEET predictions for
the Goldstone bosons \cite{PhysRevD.65.054042},
\begin{equation}
	\mu_{\pi^{\pm}}=0, \quad \mu_{K^{\pm}}=\mu_{K^0,\bar{K}^0}=\pm \frac{m_s^2-m_q^2}{2 \mu}.
	\label{eq:chem_pot}
\end{equation}
These functions are indicated by the lines in 
Fig.~\ref{fig:m30_massive_masses}. Obviously, they nicely fit the numerical 
results.

\subsection{Decay constants}

For completeness, we briefly discuss the decay constants of the 
higher-lying octet modes. Utilizing chiral Ward-Takahashi identities
one can show that either the masses or the decay constants of the 
pseudoscalar excitations vanish in the chiral limit 
(see Ref.~\cite{Kleinhaus:2007ve} for details in the context of the
present model).


\begin{figure}[hbt]
\begin{center}
 \includegraphics[width=\linewidth]{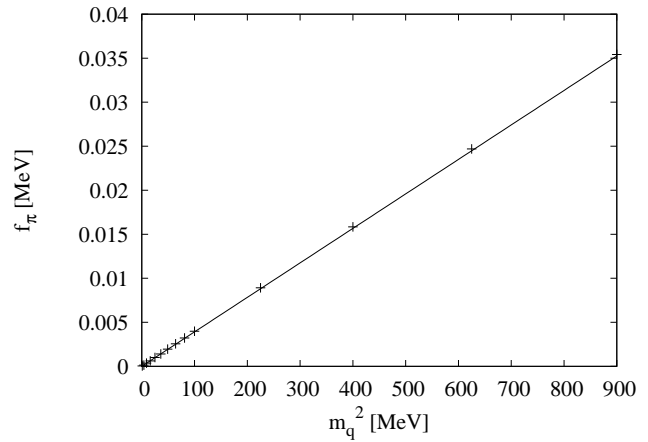}
 \caption{Decay constant of the higher-lying octet modes
as a function of the squared common quark mass $m_q$.
The points indicate the numerical results. The solid line is 
a linear fit.}
\label{fig:f_pi_massive}
\end{center}
\end{figure}


In Fig.~\ref{fig:f_pi_massive} we have plotted the decay constant
of the higher-lying octet modes as a function of the squared
common quark mass  $m_q$. 
We find that the numerical results (points) are very well described
by a straight line, meaning that $f_\pi$ behaves like $m_q^2$. 
This implies that it goes to zero in the chiral limit, as we have 
expected. 
Moreover, even for $m_q \neq 0$ it is several orders of magnitude
smaller than the decay constant of the Goldstone bosons, cf. 
Fig~\ref{fig:f_pi}.

\section{Summary \& Conclusions}
\label{summary}

We studied the properties of the pseudoscalar bosonic excitations
in the color-flavor locked phase at moderate densities within an 
NJL-type model. Extending our previous analysis \cite{Kleinhaus:2007ve},
our focus was on the hidden-flavor Goldstone bosons and on the
higher-lying pseudoscalar modes. 
Our results are consistent with the model independent predictions of the
low-energy effective theory and with predictions from axial 
Ward-Takahashi identities. 

First, we discussed the weak decay constants of the
Goldstone bosons in the chiral limit. Since the Goldstone bosons
form an $SU(3)$ octet and a singlet, there are two different decay
constants, correspondingly. For the hidden-flavor octet mesons
$\pi^0$ and $\eta$ we confirmed our previous results obtained in the
flavored sector \cite{Kleinhaus:2007ve}. In addition we calculated
the decay constant $f_{\eta'}$ of the singlet. We found that the
weak-coupling limit of $f_{\eta'}$ is correctly reproduced, whereas for
stronger couplings, i.e, for higher values of $\Delta$, we found deviations 
from this limit.

Next, we investigated the masses of the Goldstone bosons. For the 
$SU(3)$-symmetric case (equal quark masses),
the octet remains degenerate and the hidden-flavor mesons
$\pi^0$ and $\eta$ have the same mass as the flavored mesons. 
The $\eta'$, however, has a slightly higher mass. We found a
linear dependence on the quark mass in both cases. Our results are in good
agreement with the low-energy effective theory predictions if the NJL-model
values for the decay constants and the coefficient $A$ are used. 
This remains true in the case of explicit $SU(3)$ breaking via a larger
strange quark mass. 

In the last part, we studied the higher-lying pseudoscalar modes, which
appear naturally in the diagonalization procedure of the $T$-matrix for 
quark-quark scattering. We found that the octet states are bound in
most cases and are always very close to their respective threshold for 
decay into two quasiparticles. The singlet mode, on the other hand is 
unbound. 

Finally, we calculated the decay constant of the higher-lying
octet modes for equal quark masses. In agreement with axial Ward-Takahashi
identities it is very small and vanishes in the chiral limit.

\begin{acknowledgments}

We thank D.~Ebert, K.~Klimenko, D.~Nickel and M.~Oertel for valuable comments.
This work has been supported in part by the BMBF under contract 06DA123.
 
\end{acknowledgments}



\begin{thebibliography}{25}
\expandafter\ifx\csname natexlab\endcsname\relax\def\natexlab#1{#1}\fi
\expandafter\ifx\csname bibnamefont\endcsname\relax
  \def\bibnamefont#1{#1}\fi
\expandafter\ifx\csname bibfnamefont\endcsname\relax
  \def\bibfnamefont#1{#1}\fi
\expandafter\ifx\csname citenamefont\endcsname\relax
  \def\citenamefont#1{#1}\fi
\expandafter\ifx\csname url\endcsname\relax
  \def\url#1{\texttt{#1}}\fi
\expandafter\ifx\csname urlprefix\endcsname\relax\def\urlprefix{URL }\fi
\providecommand{\bibinfo}[2]{#2}
\providecommand{\eprint}[2][]{\url{#2}}

\bibitem[{\citenamefont{Kleinhaus et~al.}(2007)\citenamefont{Kleinhaus,
  Buballa, Nickel, and Oertel}}]{Kleinhaus:2007ve}
\bibinfo{author}{\bibfnamefont{V.}~\bibnamefont{Kleinhaus}},
  \bibinfo{author}{\bibfnamefont{M.}~\bibnamefont{Buballa}},
  \bibinfo{author}{\bibfnamefont{D.}~\bibnamefont{Nickel}}, \bibnamefont{and}
  \bibinfo{author}{\bibfnamefont{M.}~\bibnamefont{Oertel}},
  \bibinfo{journal}{Phys. Rev.} \textbf{\bibinfo{volume}{D76}},
  \bibinfo{pages}{074024} (\bibinfo{year}{2007}), \eprint{arXiv:0707.0632
  [hep-ph]}.

\bibitem[{\citenamefont{Alford et~al.}(1999)\citenamefont{Alford, Rajagopal,
  and Wilczek}}]{Alford:1998mk}
\bibinfo{author}{\bibfnamefont{M.~G.} \bibnamefont{Alford}},
  \bibinfo{author}{\bibfnamefont{K.}~\bibnamefont{Rajagopal}},
  \bibnamefont{and} \bibinfo{author}{\bibfnamefont{F.}~\bibnamefont{Wilczek}},
  \bibinfo{journal}{Nucl. Phys.} \textbf{\bibinfo{volume}{B537}},
  \bibinfo{pages}{443} (\bibinfo{year}{1999}), \eprint{hep-ph/9804403}.

\bibitem[{\citenamefont{Sch\"afer}(2002{\natexlab{a}})}]{Schafer:2002ty}
\bibinfo{author}{\bibfnamefont{T.}~\bibnamefont{Sch\"afer}},
  \bibinfo{journal}{Phys. Rev.} \textbf{\bibinfo{volume}{D65}},
  \bibinfo{pages}{094033} (\bibinfo{year}{2002}{\natexlab{a}}),
  \eprint{hep-ph/0201189}.

\bibitem[{\citenamefont{Rapp et~al.}(2000)\citenamefont{Rapp, Sch\"afer,
  Shuryak, and Velkovsky}}]{Rapp:1999qa}
\bibinfo{author}{\bibfnamefont{R.}~\bibnamefont{Rapp}},
  \bibinfo{author}{\bibfnamefont{T.}~\bibnamefont{Sch\"afer}},
  \bibinfo{author}{\bibfnamefont{E.~V.} \bibnamefont{Shuryak}},
  \bibnamefont{and}
  \bibinfo{author}{\bibfnamefont{M.}~\bibnamefont{Velkovsky}},
  \bibinfo{journal}{Annals Phys.} \textbf{\bibinfo{volume}{280}},
  \bibinfo{pages}{35} (\bibinfo{year}{2000}), \eprint{hep-ph/9904353}.

\bibitem[{\citenamefont{Shovkovy and Ellis}(2002)}]{Shovkovy:2002kv}
\bibinfo{author}{\bibfnamefont{I.~A.} \bibnamefont{Shovkovy}} \bibnamefont{and}
  \bibinfo{author}{\bibfnamefont{P.~J.} \bibnamefont{Ellis}},
  \bibinfo{journal}{Phys. Rev.} \textbf{\bibinfo{volume}{C66}},
  \bibinfo{pages}{015802} (\bibinfo{year}{2002}), \eprint{hep-ph/0204132}.

\bibitem[{\citenamefont{Manuel et~al.}(2005)\citenamefont{Manuel, Dobado, and
  Llanes-Estrada}}]{Manuel:2004iv}
\bibinfo{author}{\bibfnamefont{C.}~\bibnamefont{Manuel}},
  \bibinfo{author}{\bibfnamefont{A.}~\bibnamefont{Dobado}}, \bibnamefont{and}
  \bibinfo{author}{\bibfnamefont{F.~J.} \bibnamefont{Llanes-Estrada}},
  \bibinfo{journal}{JHEP} \textbf{\bibinfo{volume}{09}}, \bibinfo{pages}{076}
  (\bibinfo{year}{2005}), \eprint{hep-ph/0406058}.

\bibitem[{\citenamefont{Alford et~al.}(2007)\citenamefont{Alford, Braby, Reddy,
  and Sch\"afer}}]{Alford:2007rw}
\bibinfo{author}{\bibfnamefont{M.~G.} \bibnamefont{Alford}},
  \bibinfo{author}{\bibfnamefont{M.}~\bibnamefont{Braby}},
  \bibinfo{author}{\bibfnamefont{S.}~\bibnamefont{Reddy}}, \bibnamefont{and}
  \bibinfo{author}{\bibfnamefont{T.}~\bibnamefont{Sch\"afer}},
  \bibinfo{journal}{Phys. Rev.} \textbf{\bibinfo{volume}{C75}},
  \bibinfo{pages}{055209} (\bibinfo{year}{2007}), \eprint{nucl-th/0701067}.

\bibitem[{\citenamefont{Alford et~al.}(2008)\citenamefont{Alford, Braby, and
  Schmitt}}]{Alford:2008pb}
\bibinfo{author}{\bibfnamefont{M.~G.} \bibnamefont{Alford}},
  \bibinfo{author}{\bibfnamefont{M.}~\bibnamefont{Braby}}, \bibnamefont{and}
  \bibinfo{author}{\bibfnamefont{A.}~\bibnamefont{Schmitt}}
  (\bibinfo{year}{2008}), \eprint{0806.0285}.

\bibitem[{\citenamefont{Son and Stephanov}(2000)}]{Son:1999cm}
\bibinfo{author}{\bibfnamefont{D.~T.} \bibnamefont{Son}} \bibnamefont{and}
  \bibinfo{author}{\bibfnamefont{M.~A.} \bibnamefont{Stephanov}},
  \bibinfo{journal}{Phys. Rev.} \textbf{\bibinfo{volume}{D61}},
  \bibinfo{pages}{074012} (\bibinfo{year}{2000}), \bibinfo{note}{{\bf 62},
  059902(E) (2000)}, \eprint{hep-ph/9910491}.

\bibitem[{\citenamefont{Casalbuoni and Gatto}(1999)}]{Casalbuoni:1999wu}
\bibinfo{author}{\bibfnamefont{R.}~\bibnamefont{Casalbuoni}} \bibnamefont{and}
  \bibinfo{author}{\bibfnamefont{R.}~\bibnamefont{Gatto}},
  \bibinfo{journal}{Phys. Lett.} \textbf{\bibinfo{volume}{B464}},
  \bibinfo{pages}{111} (\bibinfo{year}{1999}), \eprint{hep-ph/9908227}.

\bibitem[{\citenamefont{Sch\"afer}(2000)}]{Schafer:2000ew}
\bibinfo{author}{\bibfnamefont{T.}~\bibnamefont{Sch\"afer}},
  \bibinfo{journal}{Phys. Rev. Lett.} \textbf{\bibinfo{volume}{85}},
  \bibinfo{pages}{5531} (\bibinfo{year}{2000}), \eprint{nucl-th/0007021}.

\bibitem[{\citenamefont{Casalbuoni et~al.}(2001)\citenamefont{Casalbuoni,
  Gatto, and Nardulli}}]{Casalbuoni:2000na}
\bibinfo{author}{\bibfnamefont{R.}~\bibnamefont{Casalbuoni}},
  \bibinfo{author}{\bibfnamefont{R.}~\bibnamefont{Gatto}}, \bibnamefont{and}
  \bibinfo{author}{\bibfnamefont{G.}~\bibnamefont{Nardulli}},
  \bibinfo{journal}{Phys. Lett.} \textbf{\bibinfo{volume}{B498}},
  \bibinfo{pages}{179} (\bibinfo{year}{2001}), \bibinfo{note}{{\bf 517}, 483(E)
  (2001)}, \eprint{hep-ph/0010321}.

\bibitem[{\citenamefont{Bedaque and Sch\"afer}(2002)}]{Bedaque:2001je}
\bibinfo{author}{\bibfnamefont{P.~F.} \bibnamefont{Bedaque}} \bibnamefont{and}
  \bibinfo{author}{\bibfnamefont{T.}~\bibnamefont{Sch\"afer}},
  \bibinfo{journal}{Nucl. Phys.} \textbf{\bibinfo{volume}{A697}},
  \bibinfo{pages}{802} (\bibinfo{year}{2002}), \eprint{hep-ph/0105150}.

\bibitem[{\citenamefont{Kaplan and Reddy}(2002)}]{PhysRevD.65.054042}
\bibinfo{author}{\bibfnamefont{D.~B.} \bibnamefont{Kaplan}} \bibnamefont{and}
  \bibinfo{author}{\bibfnamefont{S.}~\bibnamefont{Reddy}},
  \bibinfo{journal}{Phys. Rev. D} \textbf{\bibinfo{volume}{65}},
  \bibinfo{pages}{054042} (\bibinfo{year}{2002}).

\bibitem[{\citenamefont{Yamamoto et~al.}(2007)\citenamefont{Yamamoto,
  Tachibana, Hatsuda, and Baym}}]{Yamamoto:2007ah}
\bibinfo{author}{\bibfnamefont{N.}~\bibnamefont{Yamamoto}},
  \bibinfo{author}{\bibfnamefont{M.}~\bibnamefont{Tachibana}},
  \bibinfo{author}{\bibfnamefont{T.}~\bibnamefont{Hatsuda}}, \bibnamefont{and}
  \bibinfo{author}{\bibfnamefont{G.}~\bibnamefont{Baym}},
  \bibinfo{journal}{Phys. Rev.} \textbf{\bibinfo{volume}{D76}},
  \bibinfo{pages}{074001} (\bibinfo{year}{2007}), \eprint{arXiv:0704.2654
  [hep-ph]}.

\bibitem[{\citenamefont{Hong}(2000{\natexlab{a}})}]{Hong:1998tn}
\bibinfo{author}{\bibfnamefont{D.~K.} \bibnamefont{Hong}},
  \bibinfo{journal}{Phys. Lett.} \textbf{\bibinfo{volume}{B473}},
  \bibinfo{pages}{118} (\bibinfo{year}{2000}{\natexlab{a}}),
  \eprint{hep-ph/9812510}.

\bibitem[{\citenamefont{Hong}(2000{\natexlab{b}})}]{Hong:1999ru}
\bibinfo{author}{\bibfnamefont{D.~K.} \bibnamefont{Hong}},
  \bibinfo{journal}{Nucl. Phys.} \textbf{\bibinfo{volume}{B582}},
  \bibinfo{pages}{451} (\bibinfo{year}{2000}{\natexlab{b}}),
  \eprint{hep-ph/9905523}.

\bibitem[{\citenamefont{Beane et~al.}(2000)\citenamefont{Beane, Bedaque, and
  Savage}}]{Beane:2000ms}
\bibinfo{author}{\bibfnamefont{S.~R.} \bibnamefont{Beane}},
  \bibinfo{author}{\bibfnamefont{P.~F.} \bibnamefont{Bedaque}},
  \bibnamefont{and} \bibinfo{author}{\bibfnamefont{M.~J.}
  \bibnamefont{Savage}}, \bibinfo{journal}{Phys. Lett.}
  \textbf{\bibinfo{volume}{B483}}, \bibinfo{pages}{131} (\bibinfo{year}{2000}),
  \eprint{hep-ph/0002209}.

\bibitem[{\citenamefont{Nardulli}(2002)}]{Nardulli:2002ma}
\bibinfo{author}{\bibfnamefont{G.}~\bibnamefont{Nardulli}},
  \bibinfo{journal}{Riv. Nuovo Cim.} \textbf{\bibinfo{volume}{25N3}},
  \bibinfo{pages}{1} (\bibinfo{year}{2002}), \eprint{hep-ph/0202037}.

\bibitem[{\citenamefont{Ebert et~al.}(2008)\citenamefont{Ebert, Klimenko, and
  Yudichev}}]{Ebert:2007bp}
\bibinfo{author}{\bibfnamefont{D.}~\bibnamefont{Ebert}},
  \bibinfo{author}{\bibfnamefont{K.~G.} \bibnamefont{Klimenko}},
  \bibnamefont{and} \bibinfo{author}{\bibfnamefont{V.~L.}
  \bibnamefont{Yudichev}}, \bibinfo{journal}{Eur. Phys. J.}
  \textbf{\bibinfo{volume}{C53}}, \bibinfo{pages}{65} (\bibinfo{year}{2008}),
  \eprint{arXiv:0705.2666 [hep-ph]}.

\bibitem[{\citenamefont{Ebert and Klimenko}(2007)}]{Ebert:2006tc}
\bibinfo{author}{\bibfnamefont{D.}~\bibnamefont{Ebert}} \bibnamefont{and}
  \bibinfo{author}{\bibfnamefont{K.~G.} \bibnamefont{Klimenko}},
  \bibinfo{journal}{Phys. Rev.} \textbf{\bibinfo{volume}{D75}},
  \bibinfo{pages}{045005} (\bibinfo{year}{2007}), \eprint{hep-ph/0611385}.

\bibitem[{\citenamefont{Rho et~al.}(2000{\natexlab{a}})\citenamefont{Rho,
  Wirzba, and Zahed}}]{Rho:1999xf}
\bibinfo{author}{\bibfnamefont{M.}~\bibnamefont{Rho}},
  \bibinfo{author}{\bibfnamefont{A.}~\bibnamefont{Wirzba}}, \bibnamefont{and}
  \bibinfo{author}{\bibfnamefont{I.}~\bibnamefont{Zahed}},
  \bibinfo{journal}{Phys. Lett.} \textbf{\bibinfo{volume}{B473}},
  \bibinfo{pages}{126} (\bibinfo{year}{2000}{\natexlab{a}}),
  \eprint{hep-ph/9910550}.

\bibitem[{\citenamefont{Rho et~al.}(2000{\natexlab{b}})\citenamefont{Rho,
  Shuryak, Wirzba, and Zahed}}]{Rho:2000ww}
\bibinfo{author}{\bibfnamefont{M.}~\bibnamefont{Rho}},
  \bibinfo{author}{\bibfnamefont{E.~V.} \bibnamefont{Shuryak}},
  \bibinfo{author}{\bibfnamefont{A.}~\bibnamefont{Wirzba}}, \bibnamefont{and}
  \bibinfo{author}{\bibfnamefont{I.}~\bibnamefont{Zahed}},
  \bibinfo{journal}{Nucl. Phys.} \textbf{\bibinfo{volume}{A676}},
  \bibinfo{pages}{273} (\bibinfo{year}{2000}{\natexlab{b}}),
  \eprint{hep-ph/0001104}.

\bibitem[{\citenamefont{Sch\"afer}(2002{\natexlab{b}})}]{Schafer:2001za}
\bibinfo{author}{\bibfnamefont{T.}~\bibnamefont{Sch\"afer}},
  \bibinfo{journal}{Phys. Rev.} \textbf{\bibinfo{volume}{D65}},
  \bibinfo{pages}{074006} (\bibinfo{year}{2002}{\natexlab{b}}),
  \eprint{hep-ph/0109052}.

\bibitem[{\citenamefont{Abuki and Kunihiro}(2006)}]{Abuki:2005ms}
\bibinfo{author}{\bibfnamefont{H.}~\bibnamefont{Abuki}} \bibnamefont{and}
  \bibinfo{author}{\bibfnamefont{T.}~\bibnamefont{Kunihiro}},
  \bibinfo{journal}{Nucl. Phys.} \textbf{\bibinfo{volume}{A768}},
  \bibinfo{pages}{118} (\bibinfo{year}{2006}), \eprint{hep-ph/0509172}.

\end{thebibliography}
\end{document}